# Charge Transport and Inhomogeneity near the Charge Neutrality Point in Graphene

Sungjae Cho and Michael S. Fuhrer*

*Department of Physics and Center for Superconductivity Research, University of Maryland, College Park, Maryland 20742, USA*

The magnetic field-dependent longitudinal and Hall components of the resistivity $\rho_{xx}(H)$ and $\rho_{xy}(H)$ are measured in graphene on silicon dioxide substrates at temperatures from 1.6 K to room temperature. At charge densities near the charge-neutrality point $\rho_{xx}(H)$ is strongly enhanced and $\rho_{xy}(H)$ is suppressed, indicating nearly equal electron and hole contributions to the transport current. The data are inconsistent with uniformly distributed electron and hole concentrations (two-fluid model) but in excellent agreement with the recent theoretical prediction for inhomogeneously distributed electron and hole regions of equal mobility. At low temperatures and high magnetic fields $\rho_{xx}(H)$ saturates to a value $\sim h/e^2$, with Hall conductivity $<< e^2/h$, which may indicate a regime of localized $v = 2$ and $v = -2$ quantum Hall puddles.

PACS numbers: 73.63.-b, 81.05.Uw, 73.43.-f, 73.43.Qt

One of the most fascinating aspects of graphene is that the quasiparticle Hamiltonian is identical to that of massless Dirac fermions. This results in a number of interesting properties: a linear dispersion relation with a electron-hole symmetry, Fermi velocity independent of energy, absence of charge carrier mass, chirality of the electronic wavefunction (pseudospin degree of freedom), and the "Dirac point" at which the density of states vanishes linearly without the presence of an energy gap. A striking aspect of experiments is that a finite conductivity is observed in graphene for all charge densities[1], with a minimum conductivity $\sigma_{xx,min}$ on order $4e^2/h$ (but sometimes significantly smaller[2] or larger[3]) occurring near the charge-neutrality point (CNP; in the absence of disorder, the CNP and Dirac point are identical). The observation of a finite minimum conductivity has sparked significant theoretical interest. Models invoking only short-range scattering[4] give $\sigma_{xx,min} = 4e^2/\pi h$ only exactly at the CNP, and fail to reproduce the linear gate-voltage dependence of the conductivity $\sigma_{xx}(V_g)$. Other attempts[5] using the Landauer formalism also obtain $\sigma_{xx} \sim 4\ e^2/\pi h$ which depends weakly on aspect ratio, but such models are only expected to be valid in the ballistic limit for wide samples, $l < L < W$, where $l$ is the mean free path, $L$ the sample length, $W$ the sample width. Some experiments have probed this limit[2], but many do not.

In this Letter we show that the conductivity near the CNP is dominated by disorder. Distinct electron and hole puddles[6] give rise to a large magnetoresistivity with functional form consistent with theoretical work on effective media[7] and with a charge density in agreement with a self-consistent theory for Coulomb scattering in graphene[8]. At low temperatures and high magnetic fields the longitudinal resistivity $\rho_{xx}(H)$ saturates to a value $\sim h/e^2$, with Hall conductivity $\sigma_{xy}(H) \ll e^2/h$. The spatially inhomogeneous nature of the CNP[6] indicates that this "plateau"[9, 10] may in fact be due to localized $v = 2$ and $v = -2$ quantum Hall (QHE) puddles; the width in gate voltage, and temperature and field dependence of this feature are consistent with this model.

Our graphene devices are obtained by mechanical exfoliation of Kish graphite on 300 nm $SiO_2$ on silicon substrates. Graphene flakes are located using an optical microscope, and contacted with Cr/Au electrodes using electron beam lithography. Figure 1a shows an optical micrograph of a completed device; all the data in this paper are from this device.

We first characterize the density dependence of the conductivity of this device at zero and high magnetic field. Figure 1b shows the longitudinal conductivity $\sigma_{xx}$ as a function of gate voltage $V_g$. We identify the conductivity minimum at $V_g = 1.7$ V as the CNP. Away from the CNP, the conductivity increases linearly. We estimate the mobility $\mu = (1/c_g)d\sigma_{xx}/dV_g$ is 1.6 $T^{-1}$ and 2.0 $T^{-1}$ for electrons and holes respectively, where $c_g = 1.15 \times 10^{-4}$ $F/m^2$.

Figure 1c shows $\sigma_{xx}$ and the Hall conductivity $\sigma_{xy}$ as a function of gate voltage at a magnetic field of 8 T. The Hall conductivity shows the half-integer quantized plateaux that are a signature of graphene[1, 11]: $\sigma_{xy} = ve^2/h$, with $v = 4(n + 1/2)$ and $n$ an integer, $e$ the electronic charge, and $h$ Planck's constant. The plateau-like region $\sigma_{xy} \approx 0$ is also evident[9, 10].

We now discuss the magnetoresistivity $\rho_{xx}(H)$ near the CNP. Figure 2 shows $\rho_{xx}(H)$ at $V_g = 1.7$V and temperatures from 1.6 K to room temperature. At low fields the magnetoresistivity is roughly temperature independent. At higher fields the resistivity tends to saturate at a value $\sim 0.4h/e^2$ at low temperatures, and increases with no saturation for $H < 8$ T at room temperature. Figure 3 shows the gate voltage dependence of the low-field magnetoresistivity, characterized by the curvature $d^2\rho_{xx}(H)/dH^2$. The magnetoresistivity has a sharp peak at the CNP, and falls to near zero at gate voltages more than a few volts from the CNP (at $V_g = 10$V, the curvature is already three orders of magnitude lower than at the CNP).

We discuss the possible origins of the magnetoresistivity. Weak antilocalization is possible in graphene[12], and results in a positive magnetoresistivity. However, this effect should saturate at a small magnetic field scale roughly set by the coherence length squared, and should be strongly temperature dependent.

Within the Drude model, a two-dimensional conductor with a single carrier type exhibits no transverse magnetoresistivity, because the force exerted by the Hall field cancels the Lorentz force, and the drift current and resistive voltage are in the same direction. However, a conductor with electrons and holes may exhibit large transverse magnetoresistivity, because the electrons and holes may develop components of drift velocity perpendicular to the current which cancel to give zero net transverse current. It is reasonable to expect both holes and electrons to be present at zero temperature in semimetallic graphite, and at finite temperature in graphene. Such a two-fluid model has indeed been proposed to explain the gate voltage dependence of the Hall conductivity in few-layer[13] and single-layer[14] graphene. For a conductor with electrons and holes of concentrations $n$ and $p$ and of equal mobility $\mu$, $\sigma_{xx}^{n,p}(H) = \dfrac{\sigma_{xx}^{n,p}(0)}{1+(\mu H)^2}$ and $\sigma_{xy}^{n,p}(H) = \pm \dfrac{\sigma_{xx}^{n,p}(0)\mu H}{1+(\mu H)^2}$ where the positive sign is for electrons, negative for holes, and $\sigma_{xx}^{n,p}(0) = (n,p)e\mu$. Then the resistivity components are:

$$\rho_{xx}(H) = \rho_{xx}(0)\frac{1+(\mu H)^2}{1+(\alpha\mu H)^2} \qquad \rho_{xy}(H) = \alpha\mu H \rho_{xx}(H)$$

(1)

where α = (p-n)/(p+n). At the CNP, α = 0, and $\rho_{xx}(H) \propto 1 + (\mu H)^2$ and $\rho_{xy} = 0$. Far from the CNP, we expect that |α| → 1, and $\rho_{xx}(H) \approx \rho_{xx}(0)$. This model thus explains *qualitatively* the sharp peak in $\rho_{xx}(H)$ at the CNP (see Figure 3). However, it does not explain the functional form of $\rho_{xx}(H)$; Figure 4 shows $\rho_{xx}(H)$ at $T = 300$ K, open circles are the experimental data, while the dashed line is a fit to Eqn. 1 with $\mu = 1.3$ T$^{-1}$ and α = 0.38. The fit is poor; adjusting the mobility higher to fit the low-field data results in a much worse fit. Additionally, the near-absence of temperature dependence of $\rho_{xx}(H)$ is not explained by the two-fluid model; at the CNP, $n = p \approx 2.1(kT/\hbar v_F)^2$, so we expect $\rho_{xx}(0)$ and hence $\rho_{xx}(H)$ to depend quadratically on temperature. As discussed previously, another mechanism is already needed to explain the finite conductivity on order $e^2/h$ at the CNP. We put a further constraint on this mechanism: it must also explain the magnetoresistivity at the CNP.

The finite conductivity and the large magnetoresistivity at the CNP together do suggest p+n remains finite while p-n → 0. There is another scenario in which this is possible: Adam et al.[8] propose that local potential fluctuations may induce electron and hole doped regions in a nominally neutral graphene sheet. Graphene samples are characterized in terms of a single parameter, the density of Coulomb impurities $n_{imp}$, which accurately predicts the minimum conductivity, the charge density at which the CNP appears, and the field effect mobility. Within this model, the impurity density is given by $n_{imp} = (5 \times 10^{15}$ V$^{-1}$s$^{-1})\mu^{-1} \approx 2.77 \times 10^{15}$ m$^{-2}$ for our sample (using $\mu = 1.8$ T$^{-1}$, the average for electrons and holes). At the CNP, the current is carried by an effective carrier density $n^* \approx 1.08 \times 10^{15}$ m$^{-2}$, the minimum conductivity is given by $\sigma_{xx,min} = (20e^2/h)(n^*/n_{imp}) \approx 7.8e^2/h$, the CNP occurs at a gate voltage $V_{g,CNP} \approx \bar{n}e/c_g = (n_{imp}^2/4n^*)e/c_g = 2.5$ V, while the spatial charge inhomogeneity is expected to be important in a region of width $\Delta V_g = 2n^*e/c_g = 3.0$ V around the CNP. These values are in good agreement with the experimental values $\sigma_{xx,min} = 5.9e^2/h$, and $V_{g,CNP} = 1.7$ V. $\Delta V_g$ agrees well with both the width of the peak in magnetoresistivity vs. $V_g$ in Figure 3, and the width of the plateau where $\sigma_{xy} \approx 0$ in Figure 1b (~2.1 V).

We now discuss the expected magnetoresistivity for the model of Adam, et al.[8] While the general problem of magnetoresistivity in a spatially inhomogeneous conductor is complex[15], the magnetoresistivity of an inhomogeneous distribution of electrons and holes with equal mobility and equal concentrations has been solved exactly[7], and has a simple analytical form:

$$\sigma_{xx}(H) = \sigma_{xx}(0)\left(1 + (\mu H)^2\right)^{-1/2} \qquad \sigma_{xy}(H) = 0. \qquad (2)$$

Equation 2 predicts a magnetoresistivity which is linear in $H$ at high fields. Experimentally we observe a sub-linear dependence on $H$. We find however, that the room-temperature data are very well described phenomenologically by the form

$$\rho_{xx}(H) = \left(\sigma_{xx,1} + \frac{\sigma_{xx,0}}{\left(1 + (\mu H)^2\right)^{1/2}}\right)^{-1}. \qquad (3)$$

In Figure 4, we plot the experimental data (open circles) and a fit to Eqn. 3 (solid line) with $\sigma_{xx,0} = 7.1\ e^2/h$, $\sigma_{xx,1} = 0.81\ e^2/h$, and $\mu = 3.1$ T$^{-1}$. The fit is excellent. We do not yet understand the origin of the extra conductivity term in Eqn. 3, however, it is reasonable to expect deviation from Eqn. 2 for several reasons: the electron and hole concentration are not perfectly balanced, the electron and hole mobilities are not equal, and the sample geometry is far from the ideal Hall bar (some current must flow through the electrodes). However, $\sigma_{xx,0}$ is an order of magnitude larger than $\sigma_{xx,1}$, indicating that the bulk of the magnetoresistivity is described by the unusual $(1+(\mu H)^2)^{1/2}$ dependence. From the conductivity and mobility obtained from the fit to Eqn. 3 we can obtain a carrier density $n^*_{exp} = \sigma_{xx,0}/\mu e = 0.55 \times 10^{14}$ m$^{-2}$. This density is about half the predicted $n^* \approx 1.08 \times 10^{15}$ m$^{-2}$. If we use $\mu = 3.1$ T$^{-1}$ in the theory of Ref. [8] we obtain $n^* \approx 0.70 \times 10^{14}$ m$^{-2}$, in better agreement. Overall the data suggest that there is a modest enhancement of the mobility near the charge neutrality point. We also note that the calculated low-field $\rho_{xy}$ for $n = 5.5 \times 10^{14}$ m$^{-2}$ and $\mu = 3.1$ T$^{-1}$ is about ten times larger than the observed $\rho_{xy}$, in agreement with near-cancellation of electron and hole contributions to the Hall field.

At low temperatures and high magnetic fields, $\rho_{xx}(H)$ saturates to a constant value ~$0.4h/e^2$. Additionally, a plateau-like region of $\sigma_{xy} \approx 0$ is evident in $\sigma_{xy}(V_g)$. This latter feature has been interpreted as an integer quantum Hall effect (QHE) state arising either from the splitting of the valley degeneracy in the $n = 0$ Landau level (LL)[9], or due to spin splitting of the 0$^{th}$ LL resulting in counter-propagating spin polarized edge states[10]. The latter model gives rise to a dissipative QHE state, in which $\sigma_{xy}$ is only approximately quantized, and $\rho_{xx}$ is finite. Such a dissipative QHE state would also be expected in spatially inhomogeneous graphene, in which the 0$^{th}$ LL lies below or above the Fermi level in electron or hole regions respectively. The bulk then would consist of incompressible electron and hole QHE liquids, separated by regions in which the $n = 0$ LL crosses the Fermi level, i.e. fourfold degenerate edge states with counter-propagating modes. From $\mu = 3.1$ T$^{-1}$ and $n^*_{exp} = 5.5 \times 10^{14}$, we estimate the scattering time $\tau = 340$ fs, and the LL broadening $\Gamma \approx \hbar/\tau = 1.9$ meV. For $H = 8$ T the spacing between the 0$^{th}$ and 1st LL is ~100 meV, the Zeeman energy is $g\mu_B H$ = 0.9 meV assuming $g$ = 2. The average density $n^*_{exp}$ gives a LL filling factor $v$ = 2 at $H$ = 1.1 T, and 0.3 at $H$ = 8 T. Of course, the maximum density within the puddle must be greater than the average density $n^*$, and the QHE occurs over a broad range around the quantized filling factor (for example, the $v$ = +2 plateau occurs from $v$ = 1.1 - 3.3), so it is reasonable that the puddles could be in the $v$ = ±2 QHE states. Recently the imaging of electron and hole puddles in graphene was reported[16], and the puddle diameter estimated to be ~30 nm. We then expect that quantum effects should be important when the magnetic length is less than the puddle diameter, i.e. $H$ > 0.8 T, and the temperature is less than $E_F(n^*)$, i.e. T < 160 K. This is in excellent agreement with Figure 2 where significant deviation of $\rho_{xx}(H)$ from Eqn. 3 occurs at temperatures $T \leq$ 100K and $H \geq$ 0.8 T.

In conclusion, graphene exhibits pronounced magnetoresistivity very near the charge neutrality point (CNP). The magnetoresistivity is consistent with coexistence of electrons and holes near the CNP, and fits well the theoretical result[7] for spatially distinct electron and hole regions of equal mobility. At low temperatures and high magnetic fields the resistivity saturates to a value $\sim e^2/h$, and the Hall conductivity is near zero on the scale of $e^2/h$. This "$v$ = 0" Hall plateau may be the signature of local $v$ = 2 and $v$ = -2 quantum Hall regions in the sample at the CNP.

This work has been supported by the U.S. Office of Naval Research grant no. N000140610882, and NSF grant no. CCF-06-34321. Use of the nanofabrication facilities is supported by the UMD-NSF-MRSEC under grant DMR-05-20471. We are grateful to S. Adam for useful conversations.

Figure Captions

Figure 1. (a) Optical micrograph of graphene device. White vertical lines are Cr/Au electrodes, graphene is visible as slightly darker region compared to background $SiO_2$/Si substrate. (b) Longitudinal conductivity $\sigma_{xx}$ as a function of gate voltage $V_g$ at zero magnetic field and temperature of 1.6 K. (c) $\sigma_{xx}$ and Hall conductivity $\sigma_{xy}$ as a function of $V_g$ at magnetic field of 8 T and temperature of 2.3 K.

Figure 2. Longitudinal resistivity $\rho_{xx}$ as a function of magnetic field $H$ at various temperatures, and a gate voltage of 1.7 V (the point of maximum longitudinal resistivity at zero field).

Figure 3. Longitudinal conductivity $\sigma_{xx}$ (black line, left axis) and the second derivative of the longitudinal resistivity vs. magnetic field $d^2\rho_{xx}/dH^2$ at $H = 0$ (filled circles, right axis) as a function of gate voltage $V_g$ at a temperature of 1.6 K. Dotted line extrapolates between filled circles.

Figure 4. Longitudinal resistivity $\rho_{xx}$ as a function of magnetic field $H$ at a temperature of 300 K. Open circles are experimental data, dashed line is a fit to the two-fluid model (Eqn. 1 in text), and solid line is a fit to the inhomogeneous model (Eqn. 3 in text).

Figure 1

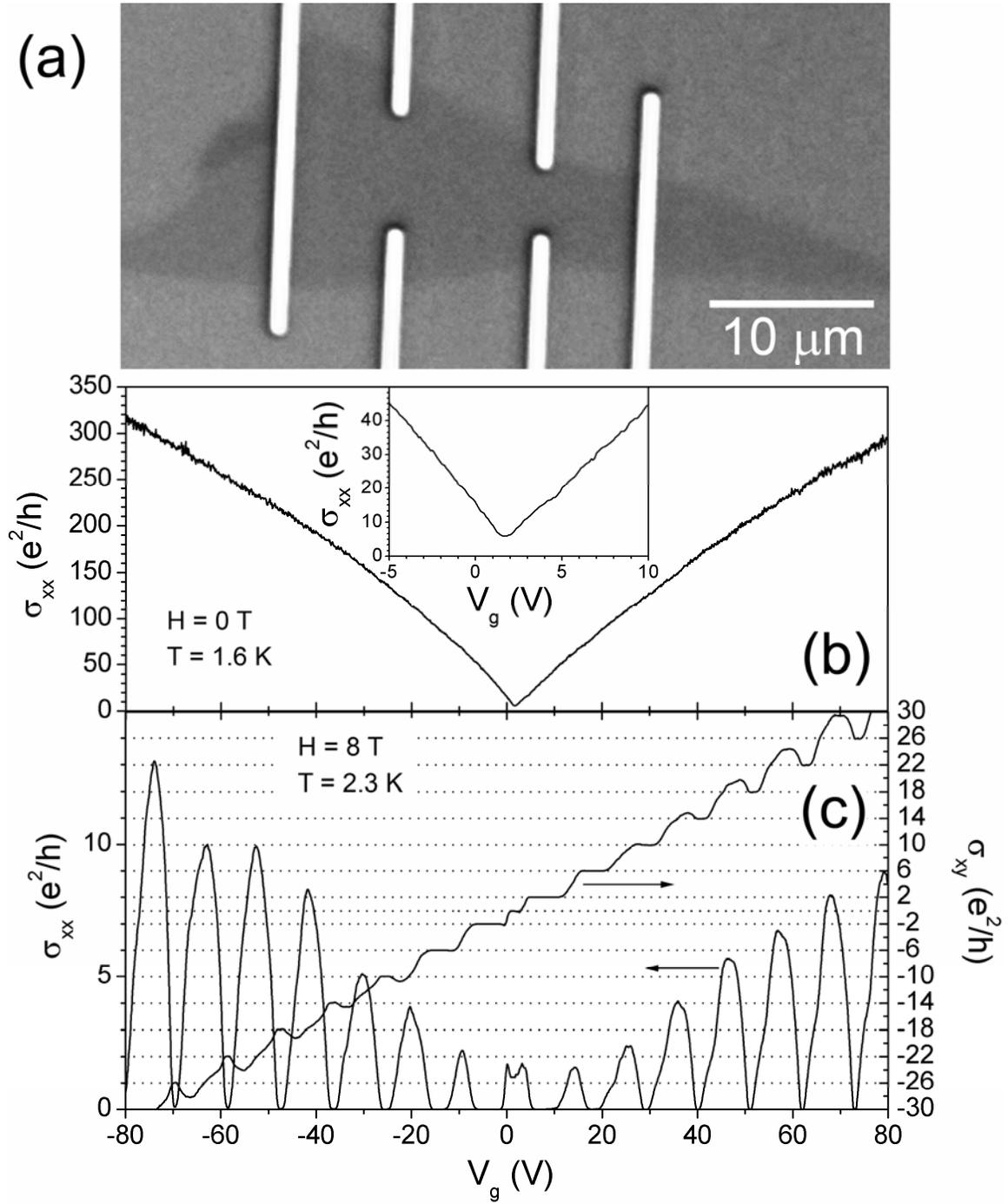

Figure 2

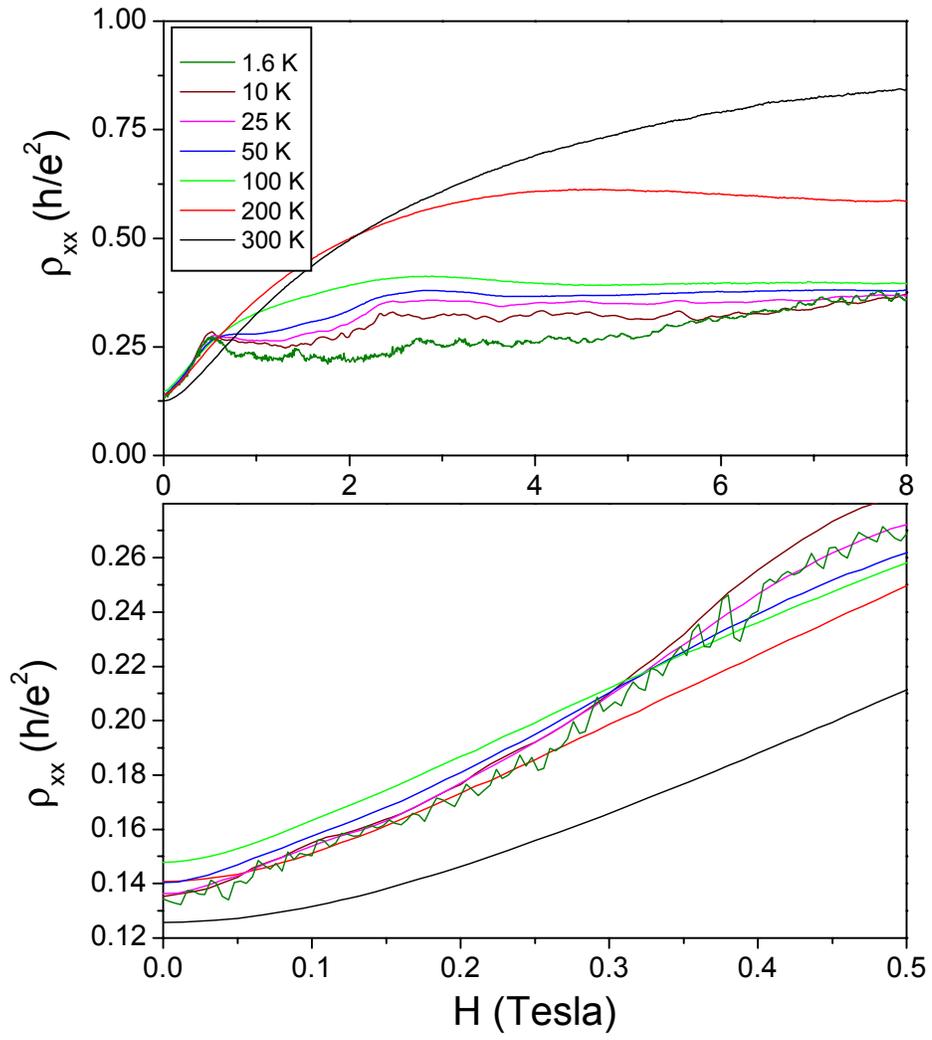

Figure 3

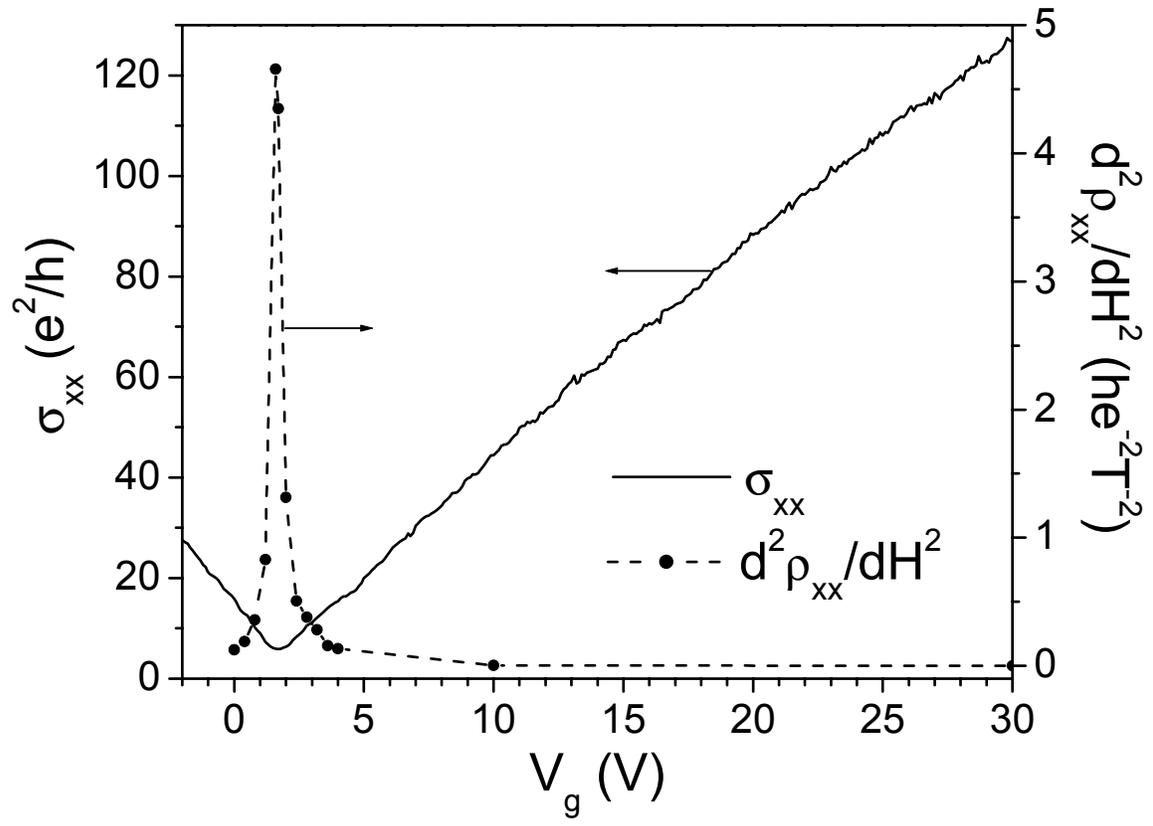

Figure 4

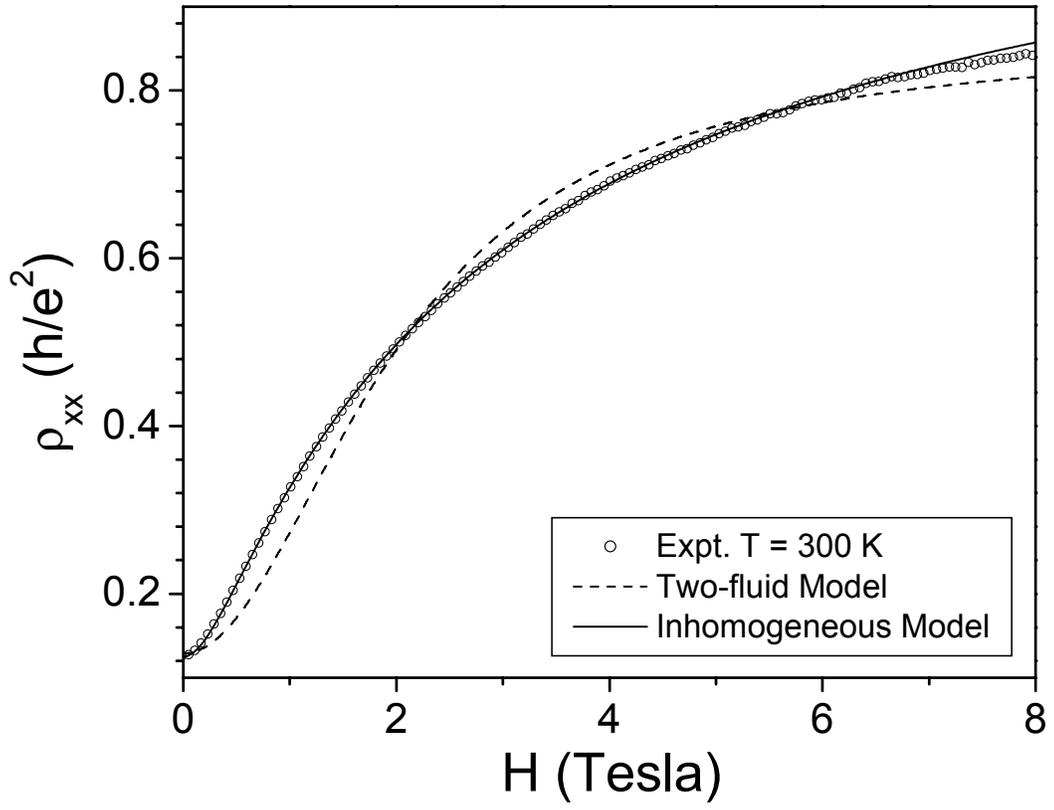